# Support for Debugging Automatically Parallelized Programs[*]

Robert Hood    Gabriele Jost

*MRJ Technology Solutions*[†]*—Numerical Aerospace Simulation Division*
*NASA Ames Research Center*
`{rhood,gjost}@nas.nasa.gov`

## Abstract

*We describe a system that simplifies the process of debugging programs produced by computer-aided parallelization tools. The system uses relative debugging techniques to compare serial and parallel executions in order to show where the computations begin to differ. If the original serial code is correct, errors due to parallelization will be isolated by the comparison.*

*One of the primary goals of the system is to minimize the effort required of the user. To that end, the debugging system uses information produced by the parallelization tool to drive the comparison process. In particular, the debugging system relies on the parallelization tool to provide information about where variables may have been modified and how arrays are distributed across multiple processes. User effort is also reduced through the use of dynamic instrumentation. This allows us to modify the program execution without changing the way the user builds the executable.*

*The use of dynamic instrumentation also permits us to compare the executions in a fine-grained fashion and only involve the debugger when a difference has been detected. This reduces the overhead of executing instrumentation.*

## 1.  Background

One of the problems facing scientific programmers on high-end computers is that as performance requirements drive up the complexity of machines, they also drive up the complexity of programming models used on them. As a consequence, debugging such codes becomes more difficult. In this paper we describe how automated debugging support can alleviate some of those problems. We begin by providing some background on the target machines and programming model that we are addressing.

### 1.1 Programming distributed memory computers

A common approach for delivering high performance in computers today is to use a distributed memory architecture. Such a computer consists of a number of processors connected together in a network. Each processor has its local memory that it can access directly. Data from other processors must be accessed via the network.

In this paper we consider the SPMD (Single Program/ Multiple Data) programming paradigm, where each processor executes the same program on a subset of the total data. Using this paradigm, computations being performed by one process will often require data calculated on another process and data has to be moved between the processes. This data movement is typically performed by explicit message passing from one processor to another using a message passing library like MPI [16] or PVM [19]. The development of a parallel program based on message passing adds a new level of complexity to the software engineering process since not only the computation, but also the explicit movement of data between processes must be specified.

Given the enormous investment made in existing scientific applications, there is a strong incentive to produce parallel versions through a conversion process rather than re-implementing from scratch. When converting a sequential program into parallel code, one way to achieve parallelism is to partition the elements of an array among the processors and have each processor update only the array elements that are assigned to it. In order to determine which updates require data from another process, the array indices have to be analyzed to detect dependencies between individual statements, iterations of a loop, and subroutine calls. The technique of dependence analysis is well understood [20] and has been implemented in compilers for code optimization. Discovering dependencies manually and then

---

[*]This work was supported through NASA contracts NAS 2-14303 and DTTS59-99-D-00437/A61812D (no kidding).

[†]The authors currently work for Computer Sciences Corp. Their address is NASA Ames Research Center, M/S T27A, Moffett Field, CA 94035.

The screen dumps in this paper have been modified from their normal screen appearance in order to aid reproducibility. The modifications include color changes as well as a resizing of some components from their defaults.

Paper appears in M. Ducassé (ed), *Proceedings of the Fourth International Workshop on Automated Debugging* (AADEBUG 2000), August 2000, Munich. COmputer Research Repository (http://www.acm.org/corr/), cs.SE/0012006; whole proceedings: cs.SE/0010035.

inserting the necessary message passing calls is a tedious and error-prone task.

## 1.2 Converting serial codes to message passing parallel codes

A straightforward way to convert a serial loop into a parallel loop based on message passing is to distribute the loop iterations among the processors. The array is logically partitioned into chunks and each processor is assigned one or more of the blocks[*]. It is then responsible for updating the array elements assigned to it. For example the Fortran loop
```
do i = 1, n
   a(i) = b(i) + 2
end do
```
could be parallelized by splitting up arrays `a` and `b` into contiguous sections. Each processor would execute:
```
do i = lower, upper
   a(i) = b(i) + 2
end do
```
where `lower` and `upper` denote the lower and upper index of the array section assigned to the processor. Now consider the loop:
```
do i = 1, n
   a(i) = b(i-1) + 2
end do
```
If array `b` is partitioned the same way as array `a`, Processor $p$ will have to access data from processor $p$-1. Therefore calls to communication routines have to be inserted. Processor $p$ has to send `b(upper)` to processor $p$+1 and receive `b(lower-1)` from processor $p$-1:
```
call send(b(upper), 1, real, p+1, ierr)
call receive(b(lower-1), 1, real, p-1, ierr)
do i = lower, upper
   a(i) = b(i-1) + 2
end do
```
The loop:
```
do i = 1, n
   a(i) = (a(i) + a(i-1)) * 0.5
end do
```
can not be executed in parallel since data from iteration `i` is dependent on data from iteration `i-1`.

There are several systems that assist in the tedious task of parallelizing codes. For example, the *CAPTools* system from the University of Greenwich [7] can take a serial program in Fortran77 and with some user guidance, turn it into a message passing parallel program. The user's role in this process is fairly modest. While the tool is analyzing the serial code, it may ask the user for additional information such as whether a subprogram that is not available for analysis modifies its parameters. After the analysis is done, the user chooses a distribution for one or more of the arrays. Then, when *CAPTools* is producing the parallel version, it may ask additional questions about the relative values of variables such as "Can N be larger than M?". The result of this process is a message passing version of the code.

## 1.3 Debugging these codes

When parallel programs are produced in this fashion, there are two types of errors that can be introduced that will cause the parallel program to behave differently from the serial version:
- bugs due to incorrect user inputs, e.g., incorrect responses to the system's queries or incorrect removal of dependences, and
- bugs due to errors in the tool itself.

For example, suppose the user incorrectly says that the statement
```
call sub(a, n)
```
does not modify array `a`. This may result in a parallelized loop where a processor uses stale values for parts of `a` instead of the up-to-date ones residing on another processor.

A programmer trying to isolate such bugs in the parallel program faces a daunting task. Not only has the serial source code been sprinkled with calls to communication libraries, but the loop structure may have undergone transformations as well. Since the parallelization tool attempted to optimize the communication patterns, the programmer must use sophisticated reasoning to determine whether a processor is in fact using a current or stale value. Figure 1 contains a small example of how serial code is transformed by *CAPTools*. The communication calls inserted there, such as `CAP_EXCHANGE`, refer to *CAPTools*-provided routines that are implemented in an appropriate way for the target machine (e.g., in MPI).

From the programmer's perspective, rather than attempting to debug the parallel program directly, a more promising approach is to determine where the parallel computation begins to differ from the serial one. This could be done by instrumenting both codes with print statements and examining the outputs, or by running two debugging sessions side-by-side. Both of these approaches have the drawback, however, that the programmer is required to deal with the machine-produced code in the parallel version.

The goal of our work is to provide automatic support for finding bugs in programs parallelized with tools. We feel that such a goal is feasible because we have:
- a reference program (serial code) for determining correct behavior, and
- mapping information from the parallelization tool that conveys how the serial program was transformed into the parallel one.

This combination permits the debugger to do side-by-side executions of the serial and parallel versions of the code. In particular the user could compare corresponding states between the two executions without being required to look at the parallel code. In the next section we will discuss possi-

---
[*]Besides the block partitioning used in the example, cyclic or block cyclic distribution of array elements is common.

```
program main
real*8 u(0:33, 0:33), v(0:33, 0:33)
call loop (u, v)
end

subroutine loop (upar, vpar)
real*8 upar (0:33, 0:33), vpar (0:33, 0:33)
integer i, j, d1,  d2
d1 = 33
d2 = 33
do i =  0, d1
  do j = 0, d2
    upar (i,j) = 0.
    vpar (i,j) = 1.
  end do
end do
do i = 1, 32
  do j= 1, 32
    upar (i,j) = upar(i,j) + 0.25 *
+                              (vpar(i-1,j   ) +
+                               vpar(i+1,j   ) +
+                               vpar(i,   j-1) +
+                               vpar(i,   j+1))
  end do
end do
return
end
```

**Original serial code.**

```
PROGRAM PARALLELmain
INTEGER CAP_LEFT,CAP_RIGHT
PARAMETER (CAP_LEFT=-1,CAP_RIGHT=-2)
REAL*8 u(0:33,0:33),v(0:33,0:33)
INTEGER CAP_BLu,CAP_BHu
COMMON /CAP_RANGE/CAP_BLu,CAP_BHu
INTEGER CAP_ICOUNT
CALL CAP_INIT
call CAP_SETUPPART(0,33,CAP_BLu,CAP_BHu)
call loop(u,v,CAP_BLu,CAP_BHu)
call dummy(u,v)
CALL CAP_FINISH()
END

subroutine loop(upar,vpar,
+           CAP_Lupar,CAP_Hupar)
integer CAP_LEFT,CAP_RIGHT
PARAMETER (CAP_LEFT=-1,CAP_RIGHT=-2)
REAL*8 upar(0:33,0:33),vpar(0:33,0:33)
integer i,j,d1,d2
integer CAP_Lupar,CAP_Hupar
COMMON /CAP_RANGE/CAP_BLu,CAP_BHu
integer CAP_BLu,CAP_BHu
integer CAP_j
d1=33
d2=33
  do i=MAX(0,CAP_Lupar),MIN(d1,CAP_Hupar),1
    do j=0,d2,1
    upar(i,j)=0.
    vpar(i,j)=1.
    enddo
  enddo
  do CAP_j=1,32,1
  CALL CAP_EXCHANGE(vpar(CAP_Hupar+1,
+                   CAP_j),
+           vpar(CAP_Lupar,CAP_j),
+           1,3,CAP_RIGHT)
  enddo
  do CAP_j=1,32,1
  CALL CAP_EXCHANGE(vpar(CAP_Lupar-1,
+                   CAP_j),
+      vpar(CAP_Hupar,CAP_j),1,3,CAP_LEFT)
  enddo
  do i=MAX(1,CAP_Lupar),MIN(32,CAP_Hupar),1
    do j=1,32,1
    upar(i,j)=upar(i,j)+0.25*(vpar(i-1,j)+
+      vpar(i+1,j)+vpar(i,j-1)+vpar(i,j+1))
    enddo
  enddo
return
END
```

**Output of *CAPTools*.**

**FIGURE 1. How *CAPTools* transforms a serial loop.**

ble approaches for automating the execution comparison process.

## 2. Relative Debugging

There are many situations in software development where it is helpful to find out how two related programs differ in behavior. One example is that of locating a bug that was introduced between versions *n* and *n+1* of a program. *Relative debugging* [1] is a technique that compares data between a program that produces correct results and one that produces faulty results to narrow down at what point discrepancies occur.

### 2.1 How relative debugging can be used to solve this problem

The technique of relative debugging is directly applicable to the situation of debugging automatically parallelized code since we can assume the existence of a sequential version that produces the correct results. Let us assume we have a sequential program $P_s$ and a parallel program $P_p$ that has been derived from $P_s$ by running its source code through a parallelization tool such as the *CAPTools* program described earlier. If $P_p$ crashes or produces wrong results, we could isolate the bugs by comparing data between $P_s$ and $P_p$. In doing such a comparison, there are several issues to address.

**What data values should be compared between the two executions?** A good starting point is a user-specified value that has been determined to be incorrect by examining the results of a previous run. The testing could be made more precise by also comparing values used to define the known incorrect one.

**When during execution should they be compared?** One possibility would be to perform the comparison immediately after any statement that could change a value of interest. This might be prohibitively expensive in execution time. Another approach is to do a comparison before and after every subroutine execution that could change the value. This effectively brackets the error location to one subroutine.

**How do we know if the values are different?** Testing equality is something that will vary from application to application. For example, in some programs scalar values may be considered "the same" if they are within some tolerance. Arrays may only be considered the same if all corresponding elements are equal. Alternatively, it may be acceptable to calculate checksums of arrays and then compare the sums.

**How do we get values from multiple address spaces to a place where they can be compared?** There are at least three approaches for this.
- One way to perform the comparison debugging would be to manually insert statements to print the array data to a file, recompile, rerun, and then inspect the printed data from the two executables. The drawbacks of this method are obvious, particularly when many processes are involved.
- Another way would be to use an enhanced debugger that controls both executables. We then have the debugger insert breakpoints, compare the data at the breakpoints, and stop when differences are detected. This approach is taken by the GUARD project [1][2][3].
- A third technique is to have the two computations establish communication, transmit and compare their data, and stop when differences are detected. This can be achieved by instrumenting the source code with routines that send or receive data and perform the comparison. This approach was used as early as 1985 at NASA Ames to debug an FFT code that had been ported to a 4 CPU Cray 2 and showed subtle intermittent problems [4]. We have subsequently successfully employed this technique when porting codes to new machine architectures.

**How is distributed data handled, as in the case where a distributed array is being compared to a serial analog?** If an element-by-element comparison is requested, the distributed array needs to be reconstituted. Thus, array distribution information is required. If checksums are being compared, each process in the parallel computation could calculate a partial checksum. Those values could then be aggregated and compared to the serial checksum.

### 2.2 The role of the parallelization tool

If a tool is used in the parallelization process, some of the questions of the previous section can be answered without user intervention. Computer-aided parallelization tools such as *CAPTools* perform three major steps:
- data dependence analysis across statements, iterations of loops, and subroutine calls,
- partitioning of array data, and
- generating necessary calls to communication library routines.

If all the information generated during these steps is gathered in a database, the following types of information is statically available:
- definition-use chains for array elements across statements and subroutines resulting from dependence analysis and
- information about which part of an array belongs to a certain processor, resulting from data partitioning.

The first item of information can be used to identify those functions and subroutines that modify a certain array and should therefore be instrumented for comparison. The second item can be used to determine how a distributed array maps to its serial analog.

### 2.3 The role of the distributed debugger

In a relative debugging system for tool-parallelized codes, the distributed debugger provides the interface for the user. For example, the user could steer the comparison activities by selecting the arrays that should be compared. In addition, the debugger controls the executions being compared, retrieves information from the parallelization database, and instruments the target programs by having appropriate function calls inserted dynamically into the executables. Finally, the presence of the debugger will permit more extensive state examination and control of execution during the steps taken to isolate the parallelization bugs.

### 3. Prototype implementation

As part of ongoing work in a debugger research project at NASA Ames, we have built a prototype relative debugging system for tool-parallelized codes. In this section we describe its implementation.

## 3.1 The foundation

Our efforts to provide automatic relative debugging support for parallelized programs build on top of three significant software systems:
- *CAPTools*, the semi-automatic parallelization tool from the University of Greenwich described earlier,
- *Dyninst*, a dynamic code adaptation toolset from the University of Maryland [8], and
- *p2d2*, a portable distributed debugger developed at NASA Ames [18].

In turn, we will briefly describe each of these systems and its role in our implementation.

*CAPTools*

Besides being used to produce the message passing program, *CAPTools* will provide vital information to the debugging system. At the heart of *CAPTools* is a dependence analysis system that examines the serial code in order to establish the safety of running loop bodies in parallel. After performing dependence analysis, it transforms the serial code to parallel form, inserting calls to communication libraries, as needed. The results of *CAPTools*'s sophisticated analysis and transformation phases are stored in a database in the file system. This makes it possible for a debugger to find out how a serial array was distributed for parallel execution and which routines modify that array [14].

*Dyninst*

We chose to use dynamic instrumentation in the project for two main reasons:
- We wanted to avoid the context switches that would result from fine-grained instrumentation being executed in the debugger. Such context switches could slow down execution by several orders of magnitude [15].
- We also wanted to minimize what was required of the user. By using dynamic instrumentation we can avoid changes to the compilation process.

The *Dyninst* project provides a portable way of inserting new code into a running program. The new code segments can be used to instrument the program in such a way that execution time does not suffer unduly. Besides inserting code segments, the *DyninstAPI* allows operations like:
- attaching to and detaching from a running process,
- inserting or removing subroutine calls from the application program,
- stopping, continuing and terminating an application program, and
- reading from and writing to areas of memory of the application program

For example, using *Dyninst* it is straightforward to patch a running program so that function execution counts are collected. While such a thing is also possible in a conventional debugger, its interpretation of each piece of instrumentation would require several context switches. This can slow down the execution time of some programs by several orders of magnitude. When done using *Dyninst*, the function counts will be collected in the address space of the program itself, and the effect on execution time is minimized.

*P2d2*

In the Portable Parallel/Distributed Debugger (*p2d2*) project at NASA Ames, we have implemented a debugger for distributed programs. One of the goals of the project is to build a debugger that is both portable across a variety of target machines and whose user interface scales to be able to debug at least 256 processes. The result of our work so far [10] is a debugger that runs on a variety of Unix-based machines and can be used on both MPI and PVM applications.

To achieve the portability goal, *p2d2* abstracted serial debugging objects and operations in a service layer. In the current implementation, this "debugger server" is in turn layered on top of *gdb*, the debugger from the Free Software Foundation [9].

In recent work we extended *p2d2* so that it could provide a global view of distributed data [11]. *P2d2* collects the local data contributions from each processor and assembles a global picture. For this process information about how the array is distributed across the processors is necessary. *P2d2* can obtain this information either from a database, such as the one produced by *CAPTools*, or by having the user provide it via a dialog box.

## 3.2 Getting the data compared

One of the efficiency concerns we had in our design was avoiding unnecessary copies of data values, especially large arrays. For example, in the case where the serial version of an array needs to be compared with its distributed analog on an element-by-element basis, we don't want to transmit both arrays to a comparison agent. Instead, we would prefer to transmit one array to the address space of the other and perform the comparison there.

Our prototype system uses two routines running in the address spaces of the target processes in order to accomplish the comparison of data from different processes.
- One routine resides in the parallelized code and sends to the serial code either the local contributions of a distributed array or the local checksum of a distributed array depending on which way of comparison is selected.
- The receiving routine is in the reference executable. It receives the local contribution from each process of the parallel executable and compares it with the corresponding data of the undistributed array. When a mismatch is detected, a special function is called to indicate that fact.

In order to insert calls to the above routines at appropriate points, we use a process that we call the *Instrumentation Server* (IS). It uses the *DyninstAPI* library [8] which allows it to control and modify executables. In a *gdb*-like manner, this program accepts commands from standard input. The most important commands are:
- `attach`: attach to a process,
- `createPoint`: create an instrumentation point in a process, and
- `insertCall`: insert a function call at an instrumentation point.

The commands take arguments such as process ID's, names of executables, routine names, and specifications of the arguments to be passed to the instrumentation functions. For example, suppose that we want to insert the function call `sub1(arg2,arg3)` at the entry point of function `sub2(arg1, arg2, arg3)` in the process whose pid is 667 and which is running executable `a.out`. The sequence of commands to the IS to do this would be:
```
attach a.out 667
createPoint 667 sub2
insertCall 667 sub1 2 3
```
Compare that with an equivalent sequence of *gdb* commands:
```
attach a.out 667
break sub2
commands
  silent
  call sub1(arg2, arg3)
  continue
end
```
In both cases the subroutine `sub1` must be linked with the executable. Using the IS, the call is actually patched into the code and will be executed on every entry to `sub2` once the target process is continued. By contrast, in the *gdb* example the target process traps on every call to `sub2`, at which point *gdb* orchestrates the call to `sub1` and the continuation of execution after that call returns. We report some preliminary findings on the comparative efficiencies of these two methods in Section 6.

Having determined how to insert the calls to the comparison routines we now need to address the question of where to insert them. The answer to this question depends very much on the desired granularity of the comparisons. For example, a comparison could be performed every time a distributed array is written to. Due to limitations in the *DyninstAPI* library, we restricted ourselves to inserting comparison routines on entry and exit of routines that modify the distributed arrays of interest. A routine showing correct values on entry but wrong values on exit is then identified as a culprit that should be further investigated.

Another issue concerning the moving and comparing of data is determining the particular comparison test to use. Our prototype implementation provides the following alternatives for checking of distributed array data:
1. The local checksums of all parts of the distributed array are added up and compared to the checksum of the undistributed array. An error is reported if the cheksum exceeds a user supplied threshold. No array distribution information is required to perform this comparison.
2. The local checksum of each part of the distributed array is compared to the corresponding partial checksum of the undistributed array. An error is reported if one of the checksums exceeds a user supplied threshold. This method requires array distribution information so that corresponding array sections can be identified.
3. An element-by-element comparison of undistributed and distributed array. This comparison, just as the previous method, requires array distribution information.

### 3.3 Putting it together

Getting the components described in Sections 3.1 and 3.2 to cooperate to solve the relative debugging problem is the job of *p2d2*. For example, it retrieves the necessary information about critical routines and array distribution information and passes it to the executables via the IS. Just as in the case of the global array viewer [11] this information is obtained by probing the *CAPTools* database mentioned earlier. As discussed in Section 3.1, *CAPTools* stores the results from the data analysis and partitioning process in a database. Given a routine name and a formal parameter name of an array, information about the partitioning can be obtained, such as which of the dimensions are partitioned and what are the bounds of the partitioned array.

From the dependence analysis phase, information about location of statements that assign to a particular array can be obtained. Given any statement in a program and an array name, *p2d2* can construct a list of all routines that might define the array value that reaches the statement. It does this by probing the *CAPTools* database.

Having retrieved information about where to instrument, *p2d2* then interacts with the IS to insert the initialization call that provides the distribution information as well as the calls that move the data between processes and perform the comparison.

### 4. How *p2d2* coordinates relative debugging

To illustrate what activities need to be coordinated, consider the following scenario from the user's perspective. After having used *CAPTools* to parallelize a program `S`, the user runs the resulting code `P` and finds it doesn't compute

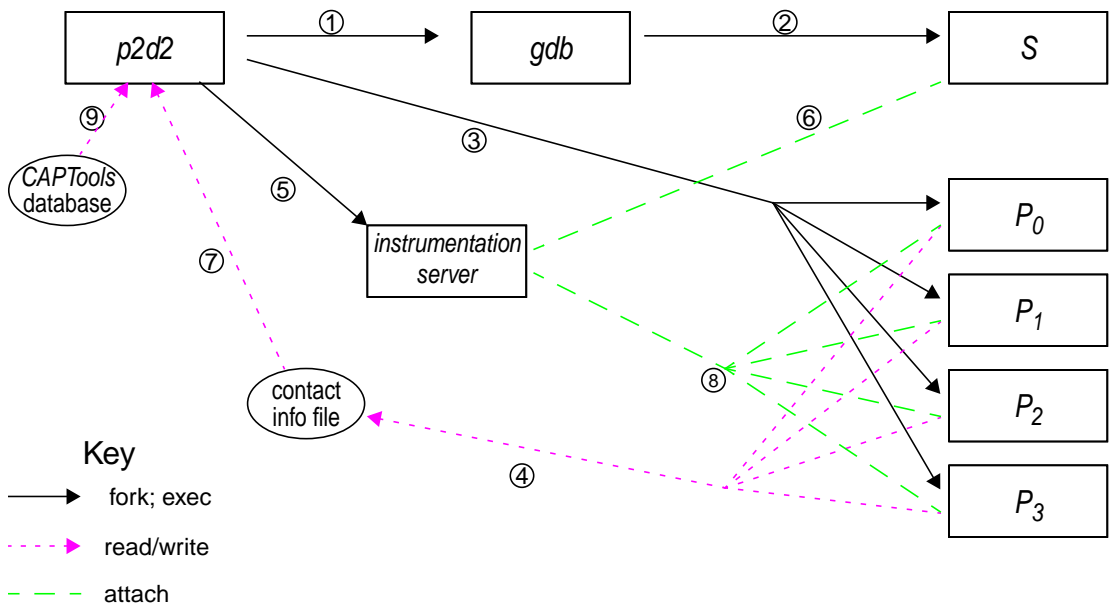

**FIGURE 2. Coordination of the comparison activities.**

the same answer. At that point, the user starts *p2d2* with the command line:

`p2d2 -R "mpirun -np 4 P" S`

which requests that *p2d2* compare the execution of "`mpirun -np 4 P`" with the execution of `S`. After *p2d2* starts up, the following sequence of events occurs. It is depicted in Figure 2.

1. *P2d2* starts a *gdb* to control execution of `S`. Then *p2d2* requests that *gdb* insert a breakpoint at entry point "`__p2d2DiffDetected`", which is the function discussed in Section 3.2 that gets called when a difference is detected.
2. The user then selects an array in *p2d2*'s source display and invokes the **Run** operation. *P2d2* issues a "`run`" request to the *gdb* controlling `S`.
3. *P2d2* issues the shell command "`mpirun -np 4 P`".
4. The four processes resulting from that command record contact information, including their process ID's, in the file system.
5. After it sees that the contact file has been created, *p2d2* starts up the IS.
6. The IS attaches to the process running `S`.
7. *P2d2* reads the parallel execution contact information from the file system. It then sends the `attach` requests to the IS.
8. The IS attaches to the four processes running `P`.
9. *P2d2* consults the *CAPTools* database and retrieves information about distributed arrays and which functions will need to be instrumented. It also provides the local name in the function of the array that needs to be monitored.

Then *p2d2* and the IS complete the instrumentation of the processes and proceed with the execution.

- The IS inserts instrumentation into the process running `S` and the four processes running `P`.
- The IS detaches from the `P` processes.
- The IS notifies *p2d2* that the instrumentation of `S` is complete. *P2d2* sends a "continue execution" request to the *gdb* controlling `S`.
- When the inserted instrumentation in `S` and `P` is executed, it establishes communication links between the serial and parallel processes.
- At the function entry and exit points that were instrumented, the parallel processes send their state information to the serial process. It compares the parallel data to its own. If there is a difference, it calls `__p2d2DiffDetected`, which causes a trap because of the breakpoint that was set there.
- When *p2d2* is notified of the trap, it determines that the cause is a difference in state between the serial and parallel executions. It then presents the information to the user.

At this point, the user has available the full power of the debugger to control and examine the serial program to isolate the problem.

```
        program jacobi                                    subroutine update (phi4, oldphi4, ...)
C       main routine                              C       Routine that updates array
        double precision phi2 (100, 100)                  ...
        double precision oldphi2(100,100)                 do j=0,nptsx+1
        ...                                                 do i=0,nptsy+1
        call setup_grid (phi2, ...)                           oldphi8(i,j) = oldphi4(i,j)
        do iter = 1, 100                                    end do
          call copyphi (oldphi2, phi2)                    end do
          call update (phi2, oldphi2, ... )               do j=1,nptsx
        end do                                              do i=1,nptsy
        call output (phi2, nptsx, nptsy)                      phi4(i,j) = 0.25*(oldphi8(i-1,j)
        return                                    #           + oldphi8(i+1,j)
        end                                       #           + oldphi8(i,j-1)
                                                  #           + oldphi8(i,j+1))
        subroutine output (phi3, ...)                       end do
C       Routine that prints the result                    end do
        ...                                               return
        do j = 0, nptsx+1                                 end
          do i = 0, nptsy+1
            phi7 (i,j) = phi3 (i,j)                       subroutine setup_grid (phi6,...)
          end do                                  C       Routine to set up the initial
        end do                                    C       grid values
        do j = 0, nptsx+1                                 ...
          write (8,*) (phi7(i,j), i = 0,                  do j=0,nptsx+1
   #                              nptsy+1)                  do i=0,nptsy+1
        end do                                                phi6(i,j) = 0.0
        return                                            end do
        end                                             end do
                                                          ...
        subroutine copyphi (oldphi5, phi5)                do j=1,nptsx
C       Routine that saves old values                       do i=1,nptsy
        ...                                                   phi6(i,j) = 1.0
        do j = 0, 43                                        end do
          do i = 0, 43                                    end do
            oldphi5(i,j) = phi5(i,j)                      return
          end do                                          end
        end do
        return
        end
```

**FIGURE 3. Jacobi program source outline.**

## 5. An example debugging session

Suppose a user is parallelizing a code with *CAPTools*. The code is a very simple Fortran implementation of a Jacobi iteration algorithm. An outline of it is shown in Figure 3.

During the parallelization process, the user examines the dependence graph with an eye for removing dependence edges that might result in unnecessary communication or might prevent parallelization altogether. While this gives the user an opportunity to improve code performance, it can also result with incorrectly behaving code.

Figure 4 shows the *CAPTools* display of the dependence edges between the two loops in routine update. After inspecting each edge, the user decides to remove the one resulting from the definition of element oldphi8(i,j) in loop 1 and the reference of oldphi8(i+1,j) in loop 2.

The user then runs the resulting code (named "par_test") and notices that the values of array phi7 printed in routine output are different from those printed by a run of the sequential version. He then invokes the *p2d2* debugger in relative debugging mode with the command:

p2d2 -R "mpirun -np 4 par_test"  serial_test where "serial_test" is the name of the serial executable.

When the *p2d2* display comes up, the user brings up a dialog box and asks for value of variable phi7 in routine output to be monitored during execution (see Figure 5).

After the user requests the start of execution, the *CAPTools* database is probed behind the scenes. The probe determines that the following arrays should be checked at entry and exit of the corresponding routines:
- oldphi5 in copyphi
- phi4 in update
- phi6 in setup_grid

After starting both executions, the routines are instrumented and execution continued. The comparison subse-

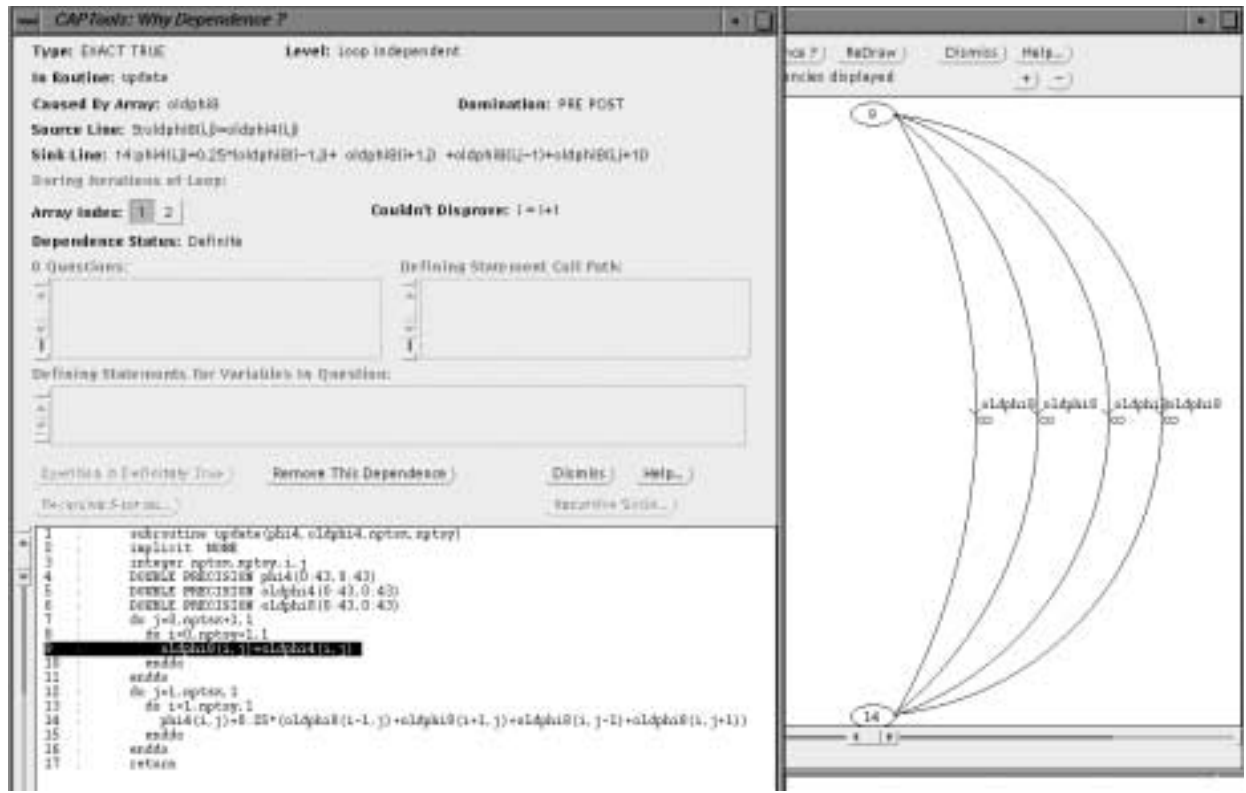

**FIGURE 4.** Examining data dependences with *CAPTools*.

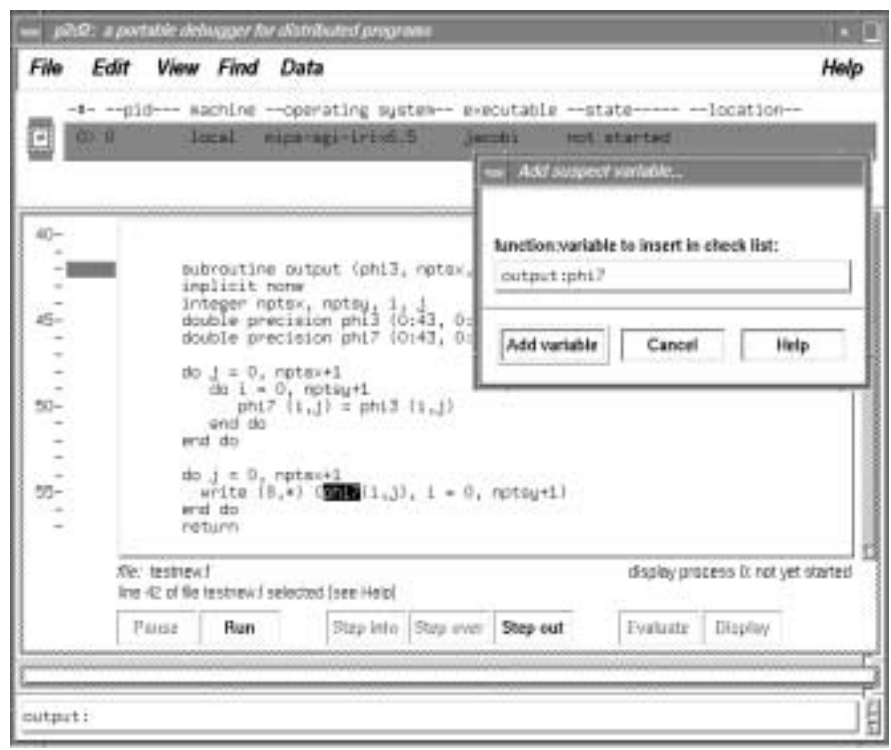

**FIGURE 5.** Preparing for a relative debugging run in *p2d2*.

quently detects a difference in `phi4` on the exit of `update`. The debugger then displays this message:

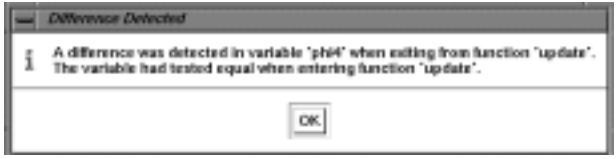

which brackets the error to execution of routine `update`. When the user inspects the parallel version of that routine:

```
    subroutine update(phi4,oldphi4, ...)
       ...
   do j=0,nptsx+1,1
     do i=MAX(0,CAP_BLphi1),
#          MIN(nptsy+1,CAP_BHphi1),1
       oldphi8(i,j)=oldphi4(i,j)
     enddo
   enddo
   CALL CAP_BEXCHANGE(
#          oldphi8(CAP_Loldphi4-1,1),
#          oldphi8(CAP_Holdphi4, 1),
#          ..., CAP_LEFT)

   do j=1,nptsx,1
     do i=MAX(1,CAP_Loldphi4),
#          MIN(nptsy,CAP_Holdphi4),1
        phi4(i,j)=0.25*(oldphi8(i-1,j)+
#                       oldphi8(i+1,j)+
#                       oldphi8(i,j-1)+
#                       oldphi8(i,j+1))
     enddo
   enddo
   return
   END
```

he sees that there is an interchange of the values of `oldphi8` to the left side to obtain the required values of `oldphi8(i-1,j)`, but there is no interchange on the right side. Therefore, stale values of `oldphi8(i+1,j)` are used to update `phi4`. The missing communication routine is due to the erroneous removal of the dependence edge earlier.

## 6. Preliminary timing results

We have performed some preliminary timing experiments to compare the impact on the execution time of the two instrumentation methods discussed in Section 3.2. The basis for our testing was a 3D Euler equation solver that ran 1000 iterations. Data comparisons were performed on entry and exit of a routine that is each called once per iteration. Each comparison was a checksum performed on an array of 40800 double precision words. In order to understand the effects of differening execution characteristics, we timed several versions of the base code, varying the following parameters:

- the size of the array being compared—a large array was 40800 double precision words and a small array was 1 double precision word,
- the amount of computation that takes place normally in the routine being instrumented, and
- the instrumentation method—no instrumentation, hand-coded instrumentation that was compiled in, IS-inserted calls to process-resident routines, or *gdb*-interpreted calls to process-resident routines.

Table 1 shows the resulting times.

While the numbers in the table give a general feeling for what is going on, we feel that a more detailed study of performance is necessary before we draw conclusions. For example, we would like to time other comparison tests such as element-wise equality, where the communication overhead will increase substantially.

Even with the limited nature of our data, it is probably safe to make these observations:

- IS-instrumented calls (based on the *DyninstAPI*) to resident functions pose very little overhead over compiled-in calls and
- *gdb*-interpreted calls are significantly more expensive.

We believe that the large array and heavy computation scenario will be the most common for our users. The preliminary numbers are not by themselves a compelling reason for our use of IS (and hence, *Dyninst*) over *gdb* for relative debugging. The relative value of the use of dynamic instrumentation increases for the other scenarios.

| Execution Characteristics | | | Instrumentation Method | | | |
|---|---|---|---|---|---|---|
| comparison test | size of array being compared | amount of computation in executable | none | by hand | IS | *gdb* |
| checksum | large | heavy | 52 | 510 | 515 | 645 |
| checksum | large | light | 13 | 440 | 443 | 560 |
| checksum | small | heavy | 52 | 52 | 52 | 190 |
| checksum | small | light | 13 | 17 | 17 | 145 |

**TABLE 1. Timings comparing the instrumentation alternatives (in seconds)**

## 7. Implementation experiences

While we see great promise in the progress to date on our goal of automatic support for debugging tool-parallelized programs, we have also observed some limitations. Many of these restrictions are imposed by the foundation software we used to build our implementation. For example, the current version of *CAPTools* does not provide information in its database about arrays distributed across more than one dimension. We are working with the implementers to ensure that this and other mapping information needed for debugging gets stored in the database.

In the case of *Dyninst*, the implementations for the platforms we tested are restricted in several ways. Perhaps the most significant is that instrumentation can currently only be placed at subroutine entry and exit. It is our understanding that eventually the package will permit instrumentation at arbitrary instructions in the code, effectively removing this restriction. In addition to this limitation, code patched in by the version of *Dyninst* we are using is unable to access function parameters in the ninth position or after. This problem is particular felt in Fortran codes, where long parameter lists are common. *Dyninst* also has a limited knowledge of the symbol table. In particular, it knows the location of global variables, but not locals or parameters.

We can get around some of the *Dyninst* symbol table limitations by using *gdb* to get that information. Unfortunately, operating system issues come up when both our *Dyninst*-based instrumentation server and *gdb* want to attach to the same process. On some systems such as Linux only one can be attached at a time. In that case, our implementation will need to coordinate attach and detach requests. Our experience on Linux shows, however, that a process cannot successfully be attached by *Dyninst* after it has been detached. For the purposes of the prototype, we restricted ourselves to an IRIX implementation where both *gdb* and our *Dyninst*-based instrumentation server could be attached at the same time.

Other issues also arise as a result of trying to debug *Dyninst*-instrumented codes. For example, when execution stops in a routine called from an instrumentation point, the runtime stack is in a state that *gdb* cannot handle—there is a return address on the stack that is outside the range that *gdb* is looking for.

In addition to limitations in the software packages used by the prototype, we should also point out some within the prototype itself. In particular, the current implementation requires the target executables to include the instrumentation routines described in Section 3.2. We plan to use dynamic linking in the future to address this restriction.

One additional limitation of our prototype is that when comparing executions we currently require that the checkpoints to be compared occur in the same order in the two runs. In the future we can address this restriction in a manner similar to Guard [2] by saving out-of-sequence checkpoints in the file system until the comparison can be made.

## 8. Related Work

*Guard* [1][2][3] is a relative debugger for parallel programs developed at the Griffith University in Brisbane Australia. In contrast to our approach, where two executables communicate data directly with each other and do the comparison, in *Guard* the debugger collects the data from the executables and does the comparison. Also, *Guard* does not aim particularly at automatically parallelized programs. Information about where to do the comparisons and what parts of the data to compare are provided by the user via the command language. To compare array data from parallel programs, the user must describe the decomposition manually using a distributed array syntax.

In other work on debugging automatically parallelized programs, Cohn [6] has investigated having the debugger provide a sequential view of an executing parallel program. While he does not use relative debugging techniques, his analysis of consistency issues between sequential and parallel executions could be useful in identifying candidate instrumentation points for making comparisons in the relative debugging approach.

The idea of using information from parallelization tools to aid in debugging has also been around for some time. For example, Hood, Kennedy, and Mellor-Crummey [12] used dependence information from a parallelizing compiler to determine which data accesses to instrument to find races in a shared-memory program execution.

## 9. Project Status and Future Work

We have built a prototype of a relative debugging system for comparing serial codes and their tool-produced parallel counterparts where array comparisons are done with either with checksums or by doing element-by-element equality tests. After the user specifies a variable and scope to be checked, the debugger uses the *CAPTools* database to determine which variables should be monitored and in which functions. We used the dynamic instrumentation tool *Dyninst* in order to minimize the overhead involved in making the comparisons. We made some preliminary timings testing the need in such an environment for dynamically inserted procedure calls versus interpreted calls. The results were inconclusive; further testing is required.

In the near future we will integrate the relative debugging features more seamlessly into *p2d2*. In particular, we would like to have debugging requests that the user makes on the serial code also be performed on the parallel version. In order to do this, we need to modify the *p2d2* user interface to support multiple computations executing simultaneously. In addition, we must get *CAPTools* to provide information about how the serial program was transformed into its parallel form. This will permit us to determine places in the code where there should be consistent state between sequential and parallel versions.

Furthermore, while *CAPTools* allows for cyclic and block-cyclic array distributions, we currently support only blockwise distributions. In the future we will address this issue.

Our approach for relative debugging of tool-parallelized distributed memory codes will also work for shared memory codes parallelized with tool support. In the near future we will extend our prototype to work with codes produced by *CAPO* [5] which is based on *CAPTools* and produces OpenMP [17] codes.

In the longer term, we would like to experiment with minimizing the time required to find differences in two computations. If a fully instrumented execution is too slow, it may be better to perform multiple partially instrumented runs, in effect doing a binary search of the computation looking for a difference. Alternatively, we could use program slice information [13], constructed with information provided by the parallelization tool, in an attempt to work backward from an incorrect value to possible definition points for the value.

Besides the relative debugging work, we would also like to experiment with other uses for dynamic instrumentation in debugging. For example, we would like to use *Dyninst* to provide fast conditional breakpoints in *p2d2*.

## 10. Conclusions

In this paper we have described a system that simplifies the process of debugging programs produced by computer-aided parallelization tools. The system uses relative debugging techniques to compare serial and parallel executions in order to show where the computations begin to differ. It uses information produced by the parallelization tool to drive the comparison process without user intervention. In addition, the use of dynamic instrumentation makes the comparisons efficient. We feel that this approach holds great promise for meeting the goal of providing automated support for isolating bugs introduced in the parallelization process.

## Acknowledgments

The authors would like to thank Henry Jin and Rob van der Wijngaart of NAS for their comments on this paper. Henry was also particularly helpful as a local expert on *CAPTools*. We also thank Steve Johnson and Peter Leggett of the University of Greenwich who were very responsive in providing routines to retrieve interprocedural information from the *CAPTools* database. Ravi Samtaney from the California Institute of Technology provided a sequential version of the RM3d code for the solution of Euler's equations in three dimensions, which we used as a test program.